\documentstyle[preprint,aps]{revtex}

\begin{document}
\title{Optical absorption for parallel cylinder arrays}
\author{P. Robles}
\address{Escuela de Ingenier\'\i a El\'ectrica, Universidad\\
Cat\'olica de Valpara\'\i so, Casilla 4059, Valpara\'\i so,\\
Chile}
\author{R. Rojas}
\address{Departamento de F\'\i sica, Universidad \\
T\'ecnica Federico Santa Mar\'\i a, Casilla 110-V, \\
Valpara\'\i so, Chile}
\author{F. Claro}
\address{Facultad de F\'\i sica, Pontificia Universidad \\
Cat\'olica de Chile, Casilla 306, Santiago 22, Chile}
\maketitle

\begin{abstract}
We study the long wavelength electromagnetic resonances of interacting
cylinder arrays. By using a normal modes expansion where the effects of
geometry and material are separated, it is shown that two parallel cylinders
with different radii have electromagnetic modes distributed symmetrically
about depolarization factor $\frac{1}{2}$. Both sets couple to longitudinal
and transverse components of the external field, but amplitudes of symmetric
depolarization factors become exchanged when considering longitudinal or
transverse polarization. We also find that amplitudes satisfy sum rules that
depend on the ratio of the cylinders radii.
\end{abstract}


\section{Introduction}

\label{sec:intro} The optical properties of ordered cylinder arrays has
become a subject of much recent interest mainly because of their potential
use as photonic crystals. \cite{yablonovitch,nicorovici,ohtaka} Theoretical
research once done for spherical particles \cite
{mcphedran,claro,rojas,olivares} has lately been applied to cylinders \cite
{radchik,reuben,proetto}. It is well known that in the long wavelength limit
the optical properties of a dilute composite of microscopic spherical
particles are well described by mean field theories such as
Claussius-Mosotti or Maxwell-Garnett. These theories are essentially based
on a dipolar approximation assuming that the particles are sufficiently far
appart so that it is possible to neglect contributions from higher order
multipoles. As particles become closer, however, this approach is no longer
valid. Several models have been presented to overcome this difficulty, among
wich the theory of normal modes has been shown to be particularly convenient
since it makes possible to expand the response of the system in terms of
optical resonance terms, where dielectric properties appear separate from
geometrical factors. \cite{berg1,berg2} The simplest system exhibiting the
effects of interactions is a pair of identical particles very close to each
other. A pair of particles of the same material and form but different size
is the simplest non-symmetric system of interacting particles \cite{olivares}%
.

Recently, a model to study arbitrary cylinder arrays made of the same
material has been constructed and applied in detail to a pair of identical
cylinders. \cite{proetto} The response of periodic arrays of identical
parallel cylinders including proximity effects is also easily treated within
that formalism. The method follows a normal modes description appropriate to
the long-wavelength limit first proposed by Bergmann \cite{berg1,berg2}, and
makes use of a basis of cylindrical harmonics solutions to Laplace's
equation. Modes are characterized by depolarization factors and strengths,
defined in such a way that an isolated cylinder exhibits a depolarization
factor $\frac{1}{2}$ and unit strength. We here follow a similar procedure
to study a pair of non-touching parallel cylinders of the same material but
different radii, for different polarizations of the external field. We show
that a difference in radii does not alter the property that depolarization
factors are symmetric about $\frac{1}{2}$, although in contrast with the
equal radii case, mixing of transverse and longitudinal modes occurs. This
property is no longer valid when the dielectric function of the cylinders is
different. We find that all normal modes are active for an external field
perpendicular to the cylinders axis, whether parallel or perpendicular to
the plane containing the axis. Modes inactive for identical cylinders have
very small strengths when their radii are not very different. Strengths of
modes with depolarization factors smaller than $\frac{1}{2}$ are exchanged
with those of depolarization modes bigger than $\frac{1}{2}$, when the
direction of the electric field changes from parallel to perpendicular.

In Sec.~\ref{sec:theory} we get the multipolar moments and the absorption
cross section for a pair of unequal cylinders. In Sec.~\ref{sec:numerical}
we present and discuss our numerical results. Finally in Sec.~\ref
{sec:conclusions} we summarize our conclusions.

\section{Theory}

\label{sec:theory} We consider a set of N parallel, infinite, uncharged
cylinders of dielectric function $\varepsilon_1$ placed in a homogeneous
medium of dielectric function $\varepsilon_2$, excited by an external
electric field whose wavelength is much longer than the cylinder radii or
separation between cylinders. The charge distribution they acquire may be
described in terms of individual multipole moments $q_{mj}$ obeying the
equations, \cite{claro}
\begin{equation}
q_{mj} = -\alpha_{mj} \;(\; V_{m} + \sum_{m^{\prime}j^{\prime}}\;
A_{mj}^{m^{\prime}j^{\prime}}\; q_{m^{\prime}j^{\prime}}\; )\; ,
\label{equat}
\end{equation}

\noindent Here $m$ is a positive or negative integer labeling the angular
momentum component along the cylinder axes, $j=1,2,...,N$ is a particle
index, $\alpha_{mj}=|m|\, a_j^{2\, |m|} (\varepsilon_1 -
\varepsilon_2)/(\varepsilon_1 + \varepsilon_2)$ are the multipolar
polarizabilities of cylinder cylinder $j$ of radius $a_j$, and $V_{m}$ are
the coefficients in the expansion of the external potential in terms of
cylindrical harmonics. The coupling coefficients are given by, \cite{proetto}

\begin{equation}
A_{mj}^{m^{\prime}j^{\prime}} = \left\{
\begin{array}{lc}
0 & \mbox{if $m \cdot m'>0$} \\
\frac{\displaystyle (-)^{m^{\prime}} (\; |m| + |m^{\prime}| - 1)!} {%
\displaystyle |m|\; !\; |m^{\prime}|\; !}\; \frac{\displaystyle e^{\;
i(m^{\prime}- m)\; \theta_{jj^{\prime}}}} {\displaystyle \rho_{jj^{%
\prime}}^{|m| +|m^{\prime}|}} & \mbox{if $m \cdot m'<0$,}
\end{array}
\right.  \label{coupling}
\end{equation}

\noindent where $(\rho_{jj^{\prime}},\theta_{jj^{\prime}})= \vec{%
\rho_{j^{\prime}}}-\vec{\rho_{j}}$ are polar coordinates in the  x-y plane
giving the relative position of cylinder $j^{\prime}$ with respect to
cylinder $j$.

As discussed in reference \cite{proetto}, if the cylinders are of the same
material one can separate in Eq. (1) terms depending on the material
susceptibility $\chi$ from those involving the geometry of the array. We
intend to follow here the same procedure and write Eq.~(\ref{equat}) as

\begin{equation}
\sum_{\mu^{\prime}}\; (\chi^{-1} \delta_{\mu\mu^{\prime}} +
H_{\mu}^{\mu^{\prime}})\; x_{\mu^{\prime}}\; = f_{\mu}\; ,  \label{normaleq}
\end{equation}

\noindent where ${\mu}$ represents the pair of indices $(m,j)$, and
\begin{eqnarray}
H_{mj}^{m^{\prime}j^{\prime}} & = & 2\pi\; (\delta_{mm^{\prime}}\;
\delta_{jj^{\prime}} + |m\; m^{\prime}|^{1/2}\; a_{j}^{|m|}\;
a_{j^{\prime}}^{|m^{\prime}|} A_{mj}^{m^{\prime}j^{\prime}})  \label{matrixh}
\\
f_{mj} & = & -2 \pi\; |m|^{1/2}\; a_{j}^{|m|}\; V_{mj}  \label{extern} \\
x_{mj} & = & \frac{q_{mj}}{|m|^{1/2}\; a_{j}^{|m|}}  \label{unknown}
\end{eqnarray}

\noindent Note the important feature that matrix $H$ depends on geometry
only and its eigenvalues $\{4 \pi n_{\mu}\}$ define the depolarization
factors $\{n_{\mu}\}$ of the array. For later convenience we write ${\bf H}=
{2 \pi ({\bf I + B)}}$, with ${\bf I}$ the unit matrix and $
B_{mj}^{m^{\prime}j^{\prime}}=|m\; m^{\prime}|^{1/2}\; a_{j}^{|m|}\;
a_{j^{\prime}}^{|m^{\prime}|} A_{mj}^{m^{\prime}j^{\prime}}$, so that the
depolarization factors $\{n_{\mu}\}$ and eigenvalues $\{\lambda_{\mu}\}$ of $%
{\bf B}$ satisfy the relation,

\begin{equation}
n_{\mu}=\frac{1}{2}(1+ \lambda_{\mu})\; .  \label{eigenrelation}
\end{equation}

\noindent Because of the property $B_{mj}^{m^{\prime}j^{\prime}}=0$ if $m
\cdot m^{\prime}> 0$ (see Eq. (2)), we write rows and columns of matrix $%
{\bf B}$ with indexes $m$ and $m^{\prime}$ following the sequence $%
1,2,\ldots,-1,-2,\ldots$, resulting in matrix ${\bf B}$ written in terms of
a new real matrix ${\bf b}$ of half its dimension, as
\begin{equation}
{\bf B = \left[
\begin{array}{cc}
0 & b \\
b & 0
\end{array}
\right]\; .}  \label{matrixB}
\end{equation}

\noindent From now on we use index $m$ and $m^{\prime}$ as positive
integers, and write the elements of matrix ${\bf b}$ as follows,
\begin{equation}
b_{mj}^{m^{\prime}j^{\prime}} = (-)^{m^{\prime}}\; \sqrt{m m^{\prime}}\;
\frac{ ( m + m^{\prime}- 1)!}{m ! m^{\prime}!}\; \frac{\displaystyle %
a_{j}^{m}\; a_{j^{\prime}}^{m^{\prime}} e^{\; i(m+m^{\prime})\;
\theta_{jj^{\prime}}}} {\displaystyle \rho_{jj^{\prime}}^{m+m^{\prime}}}
\label{bdefined}
\end{equation}

It can be shown that the eigenvalues of matrix ${\bf B}$ come in pairs with
opposite sign ${\lambda_{\mu}= \pm\ell_{\mu}}$, where ${\ell_{\mu}}$ are the
eigenvalues of ${\bf b}$ (see the appendix \ref{app:repeat}). As follows
from Eq. (7) the depolarization factors are then symmetric about the value $%
1/2$. The components of vector $x_{\mu}$ can de written in terms of the
eigenvalues of matrix ${\bf b}$ and elements of matrix ${\bf u}$ that
diagonalizes ${\bf b}$ through

\begin{equation}
{\bf u}^{-1} {\bf b}{\bf u} = {\bf \ell}\; .  \label{diagonalb}
\end{equation}

\noindent In the case of a uniform electric field $E_0$ parallel to the
plane containing the cylinder axes and perpendicular to the latter (parallel
field geometry), $x_{mj}=x_{-mj}$. Vector ${\bf x_{+}} = {x_{mj}}$ is then
given by

\begin{equation}
{\bf x_{+}} = ({\bf u}\; {\bf s}^{-1}{\bf u}^{-1}){\bf f}^{+},  \label{xpar}
\end{equation}

\noindent where

\begin{eqnarray}
s_{mj}^{m^{\prime}j^{\prime}} & = & \delta_{m m^{\prime}} \delta_{j
j^{\prime}} (\chi^{-1} + 2 \pi (1 + \ell_{mj})),  \label{s} \\
f_{mj}^{+} & = & \delta_{m1}\pi a_j E_{0}.  \label{fpar}
\end{eqnarray}

\noindent In the case of an electric field perpendicular to the plane
containing the cylinder axes (perpendicular field geometry), $x_{mj}=-x_{-mj}
$. Vector ${\bf x_{-}} = {x_{-mj}}$ is then given by,

\begin{equation}
{\bf x}_{-} = ({\bf u}\; {\bf r}^{-1}{\bf u}^{-1}){\bf f}^{-},  \label{xper}
\end{equation}

\noindent where now

\begin{eqnarray}
r_{mj}^{m^{\prime}j^{\prime}} & = & \delta_{m m^{\prime}} \delta_{j
j^{\prime}} (\chi^{-1} + 2 \pi (1 - \ell_{mj})),  \label{r} \\
f_{-mj}^{-} & = & \delta_{m1}i \pi a_j E_{0}.  \label{fper}
\end{eqnarray}

\noindent For a pair of unequal parallel cylinders with radii $a_1$ and $a_2$%
, and axis at a distance $R$ we define dimensionless parameters $%
\beta=a_2/a_1$ and $\delta=R/a_1$. Because of the properties $%
b_{mj}^{m^{\prime}j} = 0$ and $b_{m2}^{m^{\prime}1} =
(-\beta)^{m-m^{\prime}}b_{m1}^{m^{\prime}2}$, we write rows (columns) of
matrix ${\bf b}$ with particle index $j$ ($j^{\prime}$) in the sequence $1,2$
resulting in matrix ${\bf b}$ written in terms of a smaller array ${\bf g}$
as follows,
\begin{equation}
{\bf b}= \left[
\begin{array}{cc}
0 & {\bf g} \\
{\bf \overline g} & 0
\end{array}
\right]  \label{reduce}
\end{equation}

\noindent where ${\bf \overline g}$ is the transpose of ${\bf g}$, and the
elements of matrix ${\bf g}$ are given by
\begin{equation}
g_{m}^{m^{\prime}} = (-)^{m^{\prime}}\; \sqrt{m m^{\prime}}\; \frac{ ( m +
m^{\prime}- 1)!}{m ! m^{\prime}!}\; \frac{\beta^{m^{\prime}}}{\delta^{m +
m^{\prime}}}\; .  \label{gdefined}
\end{equation}

\noindent Results for a pair of cylinders with an external field in the
parallel or perpendicular configuration can be written in terms of a single
normal modes expansion as
\begin{equation}
x_{\pm mj} = \sum_{m^{\prime}j^{\prime}} \frac{C_{mj}^{m^{\prime}j^{%
\prime}}f_{\pm}} {\chi^{-1} + 4 \pi n_{\pm m^{\prime}j^{\prime}}}\; .
\label{xnormal}
\end{equation}
\noindent In this expression the upper (lower) sign corresponds to the
parallel (perpendicular) configuration, with $(m^{\prime},j^{\prime})$
labelling the excitation modes of the pair as a coupled system. It gives
just half of the multipoles; the others are obtained from the corresponding
symmetry property as given in the paragraph preceding Eqs.~(\ref{xpar}) and (%
\ref{xper}). We have defined the depolarization factors of modes,
\begin{equation}
n_{\pm m^{\prime}j^{\prime}}=\frac{1}{2}(1 \pm \ell_{m^{\prime}j^{\prime}})
\; ,  \label{nf}
\end{equation}
\noindent and coefficients corresponding to strength of modes,
\begin{equation}
C_{mj}^{m^{\prime}j^{\prime}} =
u_{mj}^{m^{\prime}j^{\prime}}(u_{1,1}^{m^{\prime}j^{\prime}}+\beta
u_{1,2}^{m^{\prime}j^{\prime}}) \; .  \label{Coeff}
\end{equation}
\noindent We have also defined
\begin{eqnarray}
f_{+} & = & f_{1,1}^{+}  \label{f+} \\
f_{-} & = & f_{-1,1}^{-} \; .  \label{f-}
\end{eqnarray}
\noindent We find that coefficients $C_{mj}^{m^{\prime}j^{\prime}}$ satisfy
the sum rules
\begin{eqnarray}
\sum_{m^{\prime}j^{\prime}} C_{m1}^{m^{\prime}j^{\prime}} & = & \delta_{m1}
\label{sum1} \\
\sum_{m^{\prime}j^{\prime}} C_{m2}^{m^{\prime}j^{\prime}} & = & \beta
\delta_{m1}\; .  \label{sum2}
\end{eqnarray}
It can be shown that, as with the original matrix ${\bf B}$, eigenvalues $%
\{\ell_{mj}\}$ of matrix ${\bf b}$ come also in pairs with opposite sign;
therefore the sets of depolarization factors $\{n_{+mj}\}$ and $\{n_{-mj}\}$
are identical (see the appendix). A given depolarization factor exhibits a
different strength depending on the direction of the external field. As seen
in the normal modes expansion given by Eq.~(\ref{xnormal}) the same strength
coefficients $C_{mj}^{m^{\prime}j^{\prime}}$ appear for depolarization
factor $n_{+m^{\prime}j^{\prime}}$ in the parallel field response and for $%
n_{-m^{\prime}j^{\prime}}$ in the perpendicular field response. Then,
strenghts corresponding to depolarization factors symmetric around value $1/2
$ are exchanged between responses corresponding to fields parallel or
perpendicular.

The magnitude of the electric dipole moment for the pair can be written as
\begin{equation}
p_{\pm} = \pi a_{1}^{2} \sum_{m^{\prime}j^{\prime}} \frac{%
C_{11}^{m^{\prime}j^{\prime}}+ \beta C_{12}^{m^{\prime}j^{\prime}}} {%
\chi^{-1} + 4 \pi n_{\pm m^{\prime}j^{\prime}}} E_{0} \;,  \label{moment}
\end{equation}
\noindent where $p_{+}$ ($p_{-}$) corresponds to a parallel (perpendicular)
field. The absorption cross section is proportional to the imaginary part of
the factor accompanying $E_{0}$ in the previous expression, a quantity we
identify as the complex effective polarizability of the pair. Thus we arrive
at a normal modes decomposition for the absorption cross section of two
parallel cylinders,
\begin{equation}
\sigma_{\pm} \sim \mbox{Imaginary} \{\sum_{m^{\prime}j^{\prime}} \frac{%
C_{11}^{m^{\prime}j^{\prime}}+ \beta C_{12}^{m^{\prime}j^{\prime}}} {%
\chi^{-1} + 4 \pi n_{\pm m^{\prime}j^{\prime}}}\}\;  \label{cross}
\end{equation}
\noindent where $\sigma_{+}$ ($\sigma_{-}$) corresponds to a parallel
(perpendicular) field. Notice that, as follows from Eq.~(\ref{Coeff}), the
numerator in the previous sum can be written as
\begin{equation}
(u_{1,1}^{m^{\prime}j^{\prime}}+\beta u_{1,2}^{m^{\prime}j^{\prime}})^2\; ,
\label{sq}
\end{equation}
\noindent and is always positive definite. According to the sum rules given
by Eqs.~(\ref{sum1}) and (\ref{sum2}), the sum of the numerators in
expansions ~(\ref{moment}) and (\ref{cross}) is $1+\beta^2$, a feature we
use in calculating the normalized strength of modes.

\section{Numerical Results}

\label{sec:numerical} We have solved numerically the eigenvalue equation for
matrix {\bf b} for the case of a pair of parallel cylinders of radii $a_1$
and $a_2$, and have calculated the depolarization factors $%
n_{m^{\prime}j^{\prime}}$ and strength coefficients $C_{mj}^{m^{\prime}j^{%
\prime}}$ according to Eqs.~(\ref{nf}) and (\ref{Coeff}). We have studied in
detail the normal modes expansion for the dipole moment of the pair as given
by Eq.~(\ref{moment}). Our most important finding is that when the radii are
not equal, modes with depolarization factors above and below the value $%
\frac{1}{2}$ mix for all orientations of the external field. This is known
not to happen when cylinders are equal.\cite{proetto}

In obtaining numerical results we use the dimensionless parameters $%
\sigma=R/(a_1 + a_2)$, that measures the center to center distance, $\beta=
a_2/a_1$, measuring how disimilar the radii are, and $\mu=(R-a_1-a_2)/a_1$,
measuring the border to border distance. In Fig.~\ref{spectrum1} we plot
modes for very close cylinders ($\sigma=1.10$) with one radius three times
bigger than the other ($\beta=3$). Modes are for the parallel configuration,
while those for the perpendicular case are obtained by mirror reflection
about depolarization factor $\frac{1}{2}$. Note that the placement of modes
are symmetric about this central value so that, as far as position is
concerned, they are indistinguishable in both configurations. The figure
shows the modes with largest amplitude, while weaker modes accumulate around
$n=\frac{1}{2}$. The sum of amplitudes is 10, as required by the sum rule
mentioned after Eq.~(\ref{sq}). Labels, included in order to match with
labelling in Fig.~\ref{radii}, are arbitrary.

Figure~\ref{radii} shows depolarization factors (a) and normalized
strengths (b) in terms of the parameter $\beta$, at fixed
$\sigma=1.1$. The sum over all 
strengths is unity, having been normalized by the 
factor $1+\beta^2$. Thus the results
at $\beta=3$ correspond to amplitudes shown in Fig.~%
\ref{spectrum1} with the normalization factor 10. In changing $\beta$ we keep constant
the parameter $\sigma$ by changing the center to center distance $R$ accordingly. Note
that modes with essentially zero strength at $\beta=1$ become important when
increasing this ratio. Labels are arbitrary and are used just to relate the
data in different figures.

Results shown in Fig.~\ref{radiia} were obtained by changing $\beta$
and the center to center separation $R$, but keeping constant the border to
border distance at the fixed value $\mu=0.4$. In this case we note that the
depolarization factors move appart with increasing $\beta$, while at
constant $\sigma$ (Fig. 2 (a)) they get closer. This is because the relation
between the edge to edge and center to center parameters is $%
\mu=(\sigma-1)(1+\beta)$, indicating that as $\beta$ goes to infinity, so does
$\mu$ if $\sigma$ is kept constant. Thus all modes should converge to the
isolated cylinder value $n=1/2$ in this case, while if $\mu$ is kept
constant, the modes converge to those of a cylinder in front of a plane at the
same distance.

\section{Conclusions}

\label{sec:conclusions} In summary, we have shown that the absorption cross
section of a pair of parallel cylinders of the same material but different
radii contains modes whose depolarization factors are symmetrically
distributed around $\frac{1}{2}$ with amplitudes depending on the direction
of the external field. When the field changes from the parallel to the
perpendicular configuration, amplitudes corresponding to symmetric
depolarization factors about the central value $1/2$ are exchanged.

\acknowledgments
This work was supported by the Fondo Nacional de Investigaci\'on Cient\'\i
fica y Tecnol\'ogica (Chile)  under grant 1990425, C\'atedra Presidencial en
Ciencias (F.C.), the Direcci\'on de Investigaci\'on of the Universidad
T\'ecnica Federico Santa Mar\'\i a and the Direcci\'on de Investigaci\'on of
the Universidad Cat\'olica de Valpara\'{\i}so.

\newpage
\appendix
\section{}

\label{app:repeat} We consider the eigenvalue equation for operator ${\bf T}$%
,
\begin{equation}
{\bf T x} = \lambda {\bf x}\; ,  \label{eigenT}
\end{equation}
in a basis of dimension $2M$. Here ${\bf T}$ has the form,
\begin{equation}
{\bf T = \left[
\begin{array}{cc}
0 & A \\
B & 0
\end{array}
\right]\; .}  \label{matrixT}
\end{equation}

\noindent where {\bf A}, {\bf B} each has dimension $M$. In writing the
eigenvectors ${\bf x}$ in terms of two smaller vectors $w$ and $v$ of
dimension $M$ the eigenvalue equation is cast into the form,
\begin{equation}
\left[
\begin{array}{cc}
0 & {\bf A} \\
{\bf B} & 0
\end{array}
\right] \left[
\begin{array}{c}
w \\
v
\end{array}
\right] = \lambda \left[
\begin{array}{c}
w \\
v
\end{array}
\right]\;
\end{equation}

\noindent or

\begin{eqnarray}
{\bf A} v & = & \lambda w  \label{AB1} \\
{\bf B} w & = & \lambda v\; .  \label{BA1}
\end{eqnarray}

\noindent From there we get the separate eigenvalue problems,
\begin{eqnarray}
{\bf (AB)} w & = & \lambda^2 w  \label{AB} \\
{\bf (BA)} v & = & \lambda^2 v\; ,  \label{BA}
\end{eqnarray}
both having the same eigenvalues $\lambda^2 $. In solving for the
corresponding eigenvectors and writing them as columns we get matrices ${\bf %
w}$ and ${\bf v}$ that diagonalize matrices ${\bf AB}$ and ${\bf BA}$. They
can be used to form a matrix ${\bf U}$ as,
\begin{equation}
{\bf U} = \frac{1}{\sqrt{2}} \left[
\begin{array}{cc}
{\bf w} & {\bf w} \\
{\bf v} & {\bf -v}
\end{array}
\right]\; ,  \label{matrixU}
\end{equation}
which diagonalizes matrix ${\bf T}$ according to the relation
\begin{equation}
{\bf U^{-1} T U = \Lambda}\; ,
\end{equation}
with matrix ${\bf \Lambda}$ given by
\begin{equation}
{\bf \Lambda} = \left[
\begin{array}{cr}
{\mbox{\boldmath $\lambda$}} & 0 \\
0 & {\mbox{\boldmath $-\lambda$}}
\end{array}
\right]\; .  \label{Lambda}
\end{equation}
Here {\boldmath $\lambda$ ($-\lambda$)} is a diagonal matrix formed by the
positive (negative) square root of the eigenvalues of matrices ${\bf AB}$ or
${\bf BA}$. Therefore the eigenvalues of matrix ${\bf T}$ come in pairs with
opposite sign. In the case of a pair of cylinders, we find the previous
feature two times. It first happens because coupling coefficients $%
A_{mj}^{m^{\prime}j^{\prime}}$ are zero if $m$ and $m^{\prime}$ have the
same sign, and then occurs also because they are zero for $j=j^{\prime}$.
The dimensionality $(4M)X(4M)$ of the original eigenvalue problem is seen to
be reduced to $MXM$ dimensions.

\begin{figure}[tbp]
\caption{Mode amplitudes for a pair of unequal cylinders under a uniform
electric field in the parallel field configuration. The raddi ratio equals $%
\beta=3$, and the separation parameter is $\sigma=1.1$. }
\label{spectrum1}
\end{figure}

\begin{figure}[tbp]
\caption{Depolarization factors (a) and normalized strengths (b)
as a function of size parameter $\beta$, for a pair of unequal
cylinders with fixed separation parameter $\sigma=1.1$ under a
uniform electric field. Labels correspond to those in Figure~1.}
\label{radii}
\end{figure}

\begin{figure}[tbp]
\caption{Same as Fig.~2, but with fixed border to border parameter $\mu=0.4
$.}
\label{radiia}
\end{figure}

\end{document}